\documentstyle[twocolumn,epsf,rotate,aps,pstricks]{revtex}
\title{Model for nucleation in GaAs homoepitaxy  
derived from first principles}
\author{P. Kratzer$^{1}$, C. G. Morgan$^{1,2}$, and M. Scheffler$^{1}$}
\address{$^{1}$Fritz-Haber-Institut der Max-Planck-Gesellschaft, 
Faradayweg 4--6, D-14195 Berlin, Germany} 
\address{$^{2}$Physics Department, Wayne State University, 
Detroit MI 48202, U.S.A. }
\def\be{\begin{equation}}
\def\ee{\end{equation}}
\def\bea{\begin{eqnarray}}
\def\eea{\end{eqnarray}}
\def\nn{\nonumber \\}
\begin{document}
\thispagestyle{empty}
\twocolumn[
\centerline{\large \bf A model for Nucleation in GaAs Homoepitaxy  
derived from First Principles} 

\medskip

\centerline{P. Kratzer$^{1}$, C. G. Morgan$^{1,2}$, and M. Scheffler$^{1}$}

\medskip

\centerline{\small \it $^{1}$Fritz-Haber-Institut der Max-Planck-Gesellschaft, 
Faradayweg 4--6, D-14195 Berlin, Germany} 
\centerline{\small \it $^{2}$Physics Department, Wayne State University, 
Detroit MI 48202, U.S.A. }

\bigskip

\narrower{\small 
The initial steps of MBE growth of GaAs on $\beta 2$-reconstructed GaAs(001) 
are investigated by performing total energy and electronic structure 
calculations using density functional theory and a repeated slab model of the 
surface. We study the 
interaction and clustering of adsorbed Ga atoms and the adsorption of 
As$_2$ molecules onto Ga atom clusters adsorbed on the surface.
The stable nuclei consist of  
bound pairs of Ga adatoms, which originate either from dimerization or from an 
indirect interaction mediated through the substrate reconstruction. 
As$_2$ adsorption is found to be strongly exothermic on sites with 
a square array of four Ga dangling bonds. 
Comparing two scenarios where the first As$_2$ gets incorporated in 
the incomplete surface layer, or alternatively in a new added layer, 
we find the first scenario to be preferable. In summary, the 
calculations suggest that nucleation of a new atomic layer is most 
likely on top of those surface regions where a partial filling of 
trenches in the surface has occurred before.}

\vspace{0.3cm}

]

\section{Introduction}
Growth of well-ordered crystals of III-V compound semiconductors 
requires the incorporation of both constituents in the correct stoichiometric
amounts. For instance, in MBE growth of GaAs, the Ga atoms and As molecules 
provided by the beam sources must be adsorbed and incorporated into the growing
surface in such a way that the surface stoichiometry is locally maintained. 
The atomistic processes behind stoichimetric growth are complex and not yet
fully understood at present.
Moreover, the substrate surfaces used for growth of arsenide compound 
semiconductors, in particular the
frequently used GaAs(001) surface, show a variety of complex surface
reconstructions. 
Under moderately arsenic-rich conditions, as are commonly used during growth, 
the GaAs(001) surface displays reconstructions with a $(2 \times 4)$ symmetry:
the $\alpha,\beta$ and $\beta2$ reconstructions, which contain 
surface As dimers as common building blocks. 
The strongly corrugated $\beta2$ reconstruction, which exposes three layers of
atoms, prevails in a wide range of growth conditions and serves as
the starting configuration for growth on the GaAs(001) substrate. 
For a well-controlled growth, it is required that this 
structure recovers after film deposition, 
at least after a short growth interruption.  However, 
it was already understood in early growth models \cite{farrel:87} that
different reconstructions may appear locally on the growing surface, acting as
metastable intermediates before a newly grown layer is completed. 
The details of these structural transformations remained unclear until
recently. 
Only with the help of detailed STM studies \cite{avery:97} has it become
possible to refine our understanding of the elementary steps of growth
\cite{itoh:98}. STM pictures taken from samples after submonolayer deposition
followed by  
a fast quench to room temperature show two major processes contributing to
growth on the $\beta2$-reconstructed surface: the filling of trenches
and the formation of small islands that later become part of the top-layer 
As dimers in the new layer. 
In the present paper, we present results of {\em ab initio} calculations 
for the atomistic steps of these two growth scenarios. 

\section{Calculations}
We performed {\em ab initio} calculations using density functional theory to 
describe exchange and correlation in the electronic many-particle system. 
A recent version \cite{PeBu96} of the generalized-gradient approximation
for the exchange-correlation functional was employed. 
All calculations were done with slabs consisting of 
seven or eight atomic layers and a $4 \times 4$ lateral unit cell.  
The bottom layer was passivated with pseudo-hydrogen atoms and kept fixed, 
while the top six or seven layers and adatoms were allowed to relax.

Our calculations use fully separable, norm-conserving 
pseudopotentials \cite{Hama89,KlBy82,FuSc98} to describe the 
electron-ion interaction, constructed from an all-electron atomic 
calculation with the GGA functional \cite{PeBu96}. 
Gonze's analysis 
\cite{GoSt91} was used to confirm that unphysical ghost states 
were not present in the separable representation.  
The wave functions were expanded in a plane wave basis \cite{BoKl97} with 
a cutoff energy of 10 Ry, and the k-space integration was 
performed with a special {\bf k}-point set, with a density equivalent 
to 64 {\bf k}-points in the Brillouin zone of the $(1 \times 1)$ 
surface unit cell.  

\section{Adsorption}
\begin{figure}
\epsfxsize=12.2cm
\begin{center}
\epsffile{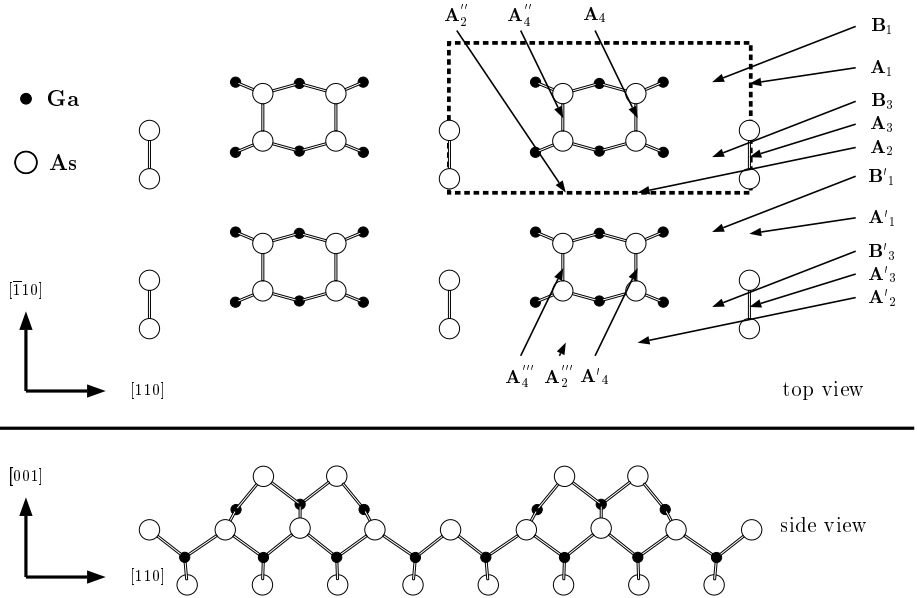}
\caption{\label{ads_sites} Adsorption sites for Ga atoms on the $\beta 2 (2
\times 4)$ reconstructed GaAs(001) surface. 
The As and Ga atoms of the substrate are displayed as white and black circles.
The unit cell of the reconstruction is marked by  the dashed rectangle.
Adsorption sites are labeled within the $4 \times 4$
cell used in the calculations.}
\end{center}
\end{figure}

Both in thermodynamic equilibrium and over a wide range of growth conditions, 
the GaAs(001) surface forms the $\beta2 (2 \times 4)$ reconstruction.
Each $(2 \times 4)$ unit cell is built up from two As dimers 
and two missing dimers in the topmost layer and a missing Ga pair in the second
layer (see Fig.~\ref{ads_sites}). 
The missing atoms give rise to trenches running in the 
$[\bar 1 1 0]$ direction separated by mountains of adjacent As dimers in the
top layer. 

\newpage
\twocolumn[
]
The As atoms in the third layer exposed in the trenches also 
form dimers. 
On the atomic level, one can imagine two principal ways in which growth can
proceed on this surface. One possibility is the nucleation of new layers on top
of the existing 'mountains'. Alternatively, the trenches could be filled up
first, either partially or completely, before nucleation of new layers 
starts afterwards in these surface regions.  

\begin{figure}
\begin{center}
\begin{tabular}{cc}
a)\epsfxsize=2.7cm
\epsffile{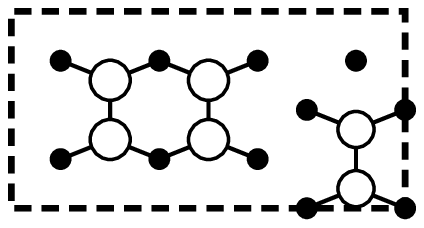} &
b)\epsfxsize=2.7cm
\epsffile{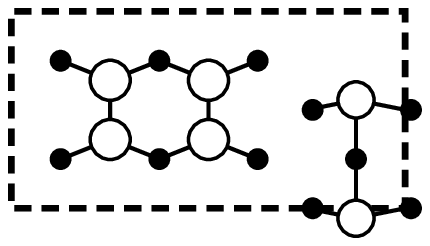}     \\
c)\epsfxsize=2.7cm
\epsffile{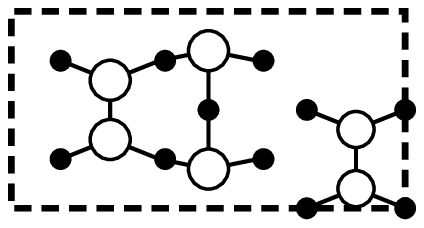}  &  
d)\epsfxsize=2.7cm
\epsffile{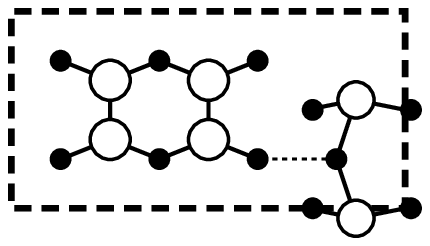}    \\
\end{tabular}
\caption{\label{Gasites} Schematic illustration of some adsorption sites of Ga
and accompaning substrate rearrangements on the GaAs(001) $\beta2$
surface: a) Ga adatom in site A$_1$ between As dimers in the
trench, b) two-fold coordinated Ga adatom in site A$_3$ splitting an As trench 
dimer,
c) two-fold coordinated Ga adatom in site A$_4$ splitting an As dimer in the top layer,
d) three-fold coordinated Ga adatom in site B$_3$ in the trench with 
additional bond to one side wall (dashed).
As atoms are shown as white circles, Ga atoms as black circles.}
\end{center}
\end{figure}

The adsorption of single gallium atoms on the GaAs(001) surface has been 
studied previously by means of density functional theory calculations 
\cite{kley:97}.
On the $\beta 2$-reconstructed surface, 
a Ga atom preferentially adsorbs between two As dimers in line with the dimer 
axis, at adsorption sites A$_1$ in the trench or in A$_2$ in the top layer, 
see Fig.~\ref{ads_sites} and Fig.~\ref{Gasites}a. 
In these calculations, the substrate atoms were allowed to relax after 
deposition of the Ga adatom, but the bonding topology of the 
substrate atoms was maintained.
A different kind of adsorption sites arises when breaking of substrate bonds is
taken into consideration. 
The Ga atom may split an As dimer in the trench (A$_3$, see 
Fig.~\ref{Gasites}b) or in the top layer
(A$_4$, see Fig.~\ref{Gasites}c) and adsorb in a two-fold coordinated site 
\cite{kley:97}.
Ga atoms in these sites are much more strongly bound than in the A$_1$ and 
A$_2$ sites. Adsorption there already constitutes
the first step to incorporation of the Ga atoms. 
Moreover, the binding energy is higher in the site A$_3$ in the trench 
than in the A$_4$ mountain site. 
As an alternative to the adsorption sites A$_1$ and A$_3$, the Ga atom 
bonding to the As atoms in the trench may tilt away from the As dimer 
axis and form an additional  bond with a Ga atom at the side wall of the 
trench (B$_1$ and B$_3$, see Fig.~\ref{Gasites}d). In this way the Ga atom reaches a 
three-fold coordination. 
Our calculations with the GGA functional show that the three-fold site 
B$_3$ obtained by splitting the As dimer is 
less strongly bound than the corresponding two-fold site A$_3$, while a 
Ga atom between As dimers experiences a stronger binding in the B$_1$ site 
than in the  A$_1$ site (see Table~\ref{clusters}).

\newpage

\section{Nucleation}
MBE growth of GaAs is usually performed by applying a flux of
arsenic molecules that exceeds the flux of Ga atoms. 
Under these conditions 
the  growth rate is mostly controlled by the 
diffusion and incorporation of the Ga atoms, while As is easily available 
everywhere on the surface. 
Therefore we focus on the interaction of Ga adatoms on the surface that 
is mainly responsible for nucleation. 

Gallium adsorption at site A$_3$ has the largest binding energy. 
Therefore, in the limit of low Ga coverage and low mobility, most of the
deposited Ga atoms will get incorporated at randomly distributed A$_3$ sites. 
At growth temperatures, however, thermally activated jumps of Ga adatoms from 
A$_3$ into neighboring sites occur frequently, with a rate of $\sim 10^5$
s$^{-1}$ \cite{footnote}.  
In this way, other adsorption sites, though being higher in energy (see
Tab.~\ref{clusters}), will become populated according to a thermal
distribution.   
If Ga adatoms in neighboring adsorption sites interact attractively, 
clusters of Ga adatoms will form which act as precursors of island growth.
Without substantial attraction, entropic effects lead to a preference for isolated single 
adsorbed Ga atoms in the small coverage limit. 

We have studied the interaction between Ga adatoms for various configurations.
The interaction energies are defined with respect to the adsorption of 
individual Ga atoms in isolated A$_3$ sites:
\be
\Delta E(A_1, \ldots A_N) = E(A_1, \ldots A_N) - N E(A_3)
\ee

\begin{table}
\begin{tabular}{l|r||l|r}
\multicolumn{2}{c||}{trench} & \multicolumn{2}{c}{top-layer} \\
\hline
singles & $\Delta E[{\rm eV}]$ & & $\Delta E[{\rm eV}]$ \\ 
\hline
A$_1$ &  0.55  & A$_2$ & 0.60 \\ 
B$_1$ &  0.20  & A$_4$ & 0.20 \\ 
B$_3$ & 0.15 & & \\ 
\hline
\multicolumn{4}{l}{pairs}\\
\hline
A$_3$A$^{'}_3$ & 0.05 & A$_4$A$^{'}_4$ & 0.20 \\
A$_3$A$_1$ &  0.90 & A$_4$A$_2$ & 0.20 \\ 
B$_3$B$^{'}_3$& 0.25 & A$_4$A$^{''}_4$ & $-$1.15 \\ 
B$_3$B$_1$ & $-$0.65 \\ 
\hline
\multicolumn{4}{l}{triples} \\
\hline
A$_3$A$_1$A$^{'}_3$ & 0.00 & A$_4$A$_2$A$^{'}_4$ & 0.10 \\
B$_3$B$_1$B$^{'}_3$ & $-$1.00 & A$_4$A$_2$A$^{''}_4$ &  $-$0.65 \\ 
\hline
\multicolumn{4}{l}{quadruples} \\
\hline
& & A$_4$A$^{''}_4$A$^{'}_4$A$^{'''}_4$ & $-$2.20 \\ 
& & A$_4$A$_2$A$^{''}_4$A$^{''}_2$ & $-$0.35 \\ 
\end{tabular}

\caption{\label{clusters} Interaction energy of clusters of Ga adatoms 
in the trenches (left) and in the top layer (right) of the $\beta2$ reconstruction. 
All numbers are given relative to isolated Ga adatoms in A$_3$ sites, with 
negative numbers indicating attractive interaction, with accuracy 50 meV.}
\end{table}

\begin{figure}[htb]
\begin{center}
\epsfxsize=7.5cm
\epsffile{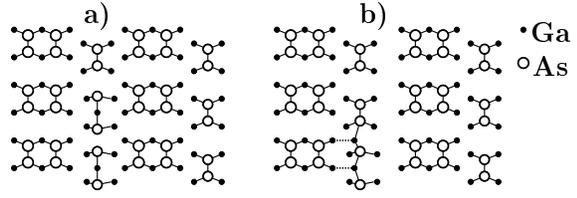}

\vspace{0.3cm}

\caption{\label{nucle} 'Trench' adsorption geometries (cf. left column in
Tab.~\protect\ref{clusters}) for Ga atoms in a trench:
a) two Ga atoms in sites A$_3$A$^{'}_3$ splitting two As dimers,
b) in adjacent B$_3$B$_1$ sites, thereby splitting only one As dimer.}   
\end{center}
\end{figure}

The left column in Table~\ref{clusters} shows the interaction energy for Ga atoms adsorbed in the
trench.
The Ga atoms alternatingly occupy adjacent sites of the type A$_1$ and A$_3$, 
or B$_1$ and B$_3$. 
In this way, Ga--As--Ga \ldots chains in 
$[\bar 1 10]$ direction are formed from the As dimers in the trenches.  
From the calculated energetics, we derive the following growth scenario: 
When a diffusing Ga adatom approaches another Ga adatom in an 
A$_3$ site, they can stabilize each other by forming a B$_3$B$_1$ pair.
This process results in a local 
$\alpha$ reconstruction in one particular $(2 \times 4)$ unit cell. 
Although the Ga atoms do not form bonds with each other,
such a pair is bound by 0.6 eV. Attachment of a third Ga adatom increases 
the interaction energy by only 0.4 eV, to 1.0 eV. 
The unexpected stability of a Ga pair is due to the fact that an As dimer bond
is restored when the diffusing Ga atom moves from the next-nearest neighbor
site to the nearest-neighbor site  with respect to a Ga atom adsorbed in the 
A$_3$ site (see Fig. \ref{nucle} for illustration). 

Since the clustering of only two Ga atoms already results in a large 
increase in the binding energy, we expect the rate of formation of 
such pairs from 
a lattice gas of single diffusing Ga adatoms to exceed the rate of break-up
of pairs even for a moderate supersaturation of Ga on the surface. Thus our 
calculations demonstrate that the Ga adatom pair in the
trench acts as a stable nucleus in the sense of nucleation theory. 

Next we consider the alternative scenario, nucleation of a new layer without 
previous trench filling.  
The calculated results are collected in the right column of
Table~\ref{clusters}. 
In general, Ga adsorption in the top-layer As dimers is energetically less favorable 
than adsorption in the trench dimers.  
Therefore population of the top-layer sites is considerably smaller for
low coverages, and only increases when the trench dimer sites are mostly
occupied. 
However, when two Ga atoms adsorb in neighboring parallel As dimers, 
they reach a stable configuration due to the formation of a Ga dimer 
(A$_4$A$^{''}_4$ in Tab. \ref{clusters}).
The decay of the Ga dimer into two single Ga adatoms in A$_3$ sites is an
endothermic process.  
This relative stability  is maintained for those larger clusters that allow for
formation of Ga dimers oriented perpendicularly to the previous 
As dimers in the layer below, i.e. the A$_4$A$_2$A$^{''}_4$ and 
A$_4$A$^{''}_4$A$^{'}_4$A$^{'''}_4$ clusters. 
We note that the latter cluster has  twice the binding energy of a
single Ga dimer, A$_4$A$^{''}_4$. Thus there is no extra attractive interaction
between Ga dimers. If larger islands of Ga addimers form on surface regions 
where the trenches are already filled up with Ga atoms, this is not due to
an attractive interaction. However, such Ga islands may form accidentally and
survive due to a limited mobility of the dimerized  
Ga atoms constituting these clusters. 

After a cluster of Ga adatoms has formed in the trench, any
further Ga adatoms that are deposited on the adjacent mountains will
be more likely to remain on the mountains long enough to
form Ga dimers there, since they would have to move further
to find empty trench sites to occupy.  This suggests
that the formation of a Ga cluster in the trench acts to
promote the subsequent formation of Ga dimer pairs on the
adjacent mountains.

In contrast to the situation in the trench, the formation of  Ga--As--Ga \ldots
chains in $[\bar 1 10]$ direction  by splitting top-layer As dimers 
is associated with only a minor gain in binding energy
(clusters A$_4$A$_2$, A$_4$A$_2$A$^{'}_4$).  
Relative to individual Ga atoms sitting in top-layer As dimers (A$_4$ sites), 
forming a chain containing two (three) Ga atoms gives a binding energy of
0.2~eV (0.5~eV). However, these structures are unstable against decay into 
individual Ga adatoms in A$_3$ trench sites, and therefore should have little
relevance for growth.   

\section{The role of arsenic}
Up to now, we have not considered the possibility of enhanced stability of
the above structures due to adsorption of arsenic.   
As sources of arsenic in MBE growth, both As$_2$ and As$_4$ molecular beams are
in use. For the issue of enhanced stability due to arsenic, 
it is sufficient to consider the simpler case of As$_2$
adsorption. While As$_4$ is believed to split into fragments upon adsorption,
the As$_2$ molecule can become incorporated into the $\beta2 (2 \times 4)$
structure without dissociation. 
{\em Ab initio}  calculations \cite{MoKr:98,shiraishi:96a} 
show that the 
binding energy of chemisorbed As$_2$ depends very much on the local 
environment. On an ideal $\beta2$-reconstructed
surface, an arsenic addimer is bound to the top-layer As dimers by 
1.65 eV \cite{MoKr:98}. 
It can be shown that As ad-dimers bound to Ga atoms 
will stay permanently adsorbed even at standard growth temperatures, 
while As dimers that bind to As atoms or to
only one Ga atom are more weakly bound and will either desorb or react 
with diffusing Ga atoms \cite{MoKr:98}.  
This is consistent with the experimental observation that arsenic incorporation
at these temperatures only proceeds when the surface is simultaneously exposed
to a Ga beam providing excess Ga adatoms on the surface \cite{foxon:77}.

\begin{figure}[b]
\leavevmode
a) \epsfxsize=7.5cm
\epsffile{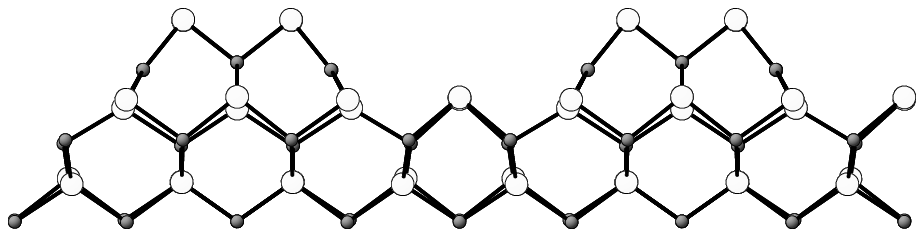}

\vspace{0.3cm}

b) \epsfxsize=7.5cm
\epsffile{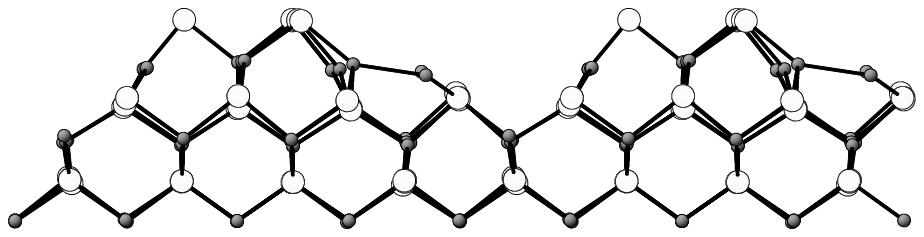}

\vspace{0.3cm}

c) \epsfxsize=7.5cm
\epsffile{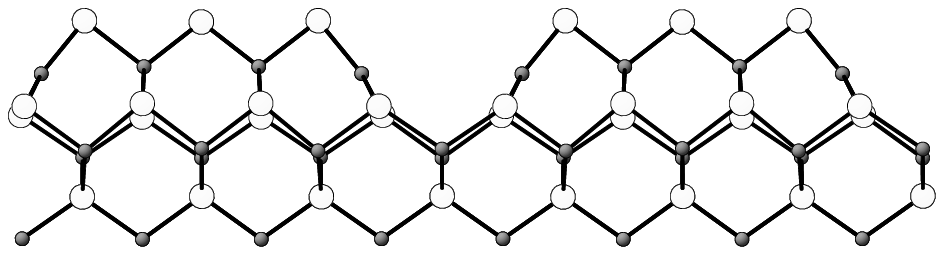}

\vspace{0.3cm}

\caption{\label{trench-scenario} Growth scenario for filling of the trenches on
the clean $\beta 2$ surface (a), via formation of Ga atom pairs B$_3$B$_1$ (b),
followed by As$_2$ adsorption that leads to a local $\beta$ reconstruction (c).
 The pictures represent show part of the slab in side view, with relaxed atomic geometries 
 as obtained from the calculations.}
\end{figure}

We expect that enhanced stability due to As$_2$ adsorption 
is most relevant for those  structures containing L-shaped patches of three Ga 
atoms or rectangular
patches of four Ga atoms in adjacent sites. In these local environments, an adsorbing As$_2$ 
molecule can attach to the surface by building up three or four As--Ga
backbonds.
For the nucleation scenario in the trenches, such a situation occurs 
already for a single Ga atom adsorbed in the trench, for instance in the A$_3$
site. The As$_2$ forms one backbond to this Ga atom, while simulataneously 
backbonds with two other dangling orbitals of Ga atoms at the sidewall of 
the trench are established. The binding energy of an As$_2$ molecule relative 
to the gas phase is 1.9 eV at this site, too low to make it permanently
adsorbed at frequently used growth temperatures.  
When two Ga atoms are on adjacent sites in the trench, like in the
local $\alpha$ structure (cluster B$_3$B$_1$, see also
Fig.~\ref{trench-scenario}b), the adsorbing 
As$_2$ can establish four backbonds and will transform this structure to the  
$\beta$ reconstruction, that contains three parallel As dimers (see Fig.~\ref{trench-scenario}c).
The binding energy of As$_2$ is 2.4 eV in this environment\cite{MoKr:98}. 
Further As$_2$ adsorption on the local $\beta$ reconstruction, which
would lead to a complete filling of the trench, has been found to be
energetically unfavorable. 
We find a binding energy of only 0.9 eV
for an As dimer filling in the fourth and last As dimer site in the
top layer of the $(2 \times 4)$ unit cell.  Since this binding energy
is much lower than the binding energies for an As addimer in other
sites which we have found to be ultimately unstable at standard growth
temperatures \cite{MoKr:98},
we conclude that the filling in of the fourth dimer site
on the local $\beta$ surface should not
play a major role as an intermediate configuration
during standard growth of the $\beta2$ surface.  
Moreover, we find it to be more favorable for an As$_2$ molecule 
adsorbing on a local $\beta$ structure to attach itself onto the As top layer,
with its axis oriented perpendicularly to the existing As dimers.  
We expect
that complete filling of the trench does not occur until nucleation of
the new mountains of the next layer up changes the structure so that it
is no longer locally the $\beta$ structure.

We note that it may not be easy to distinguish experimentally 
between a three-dimer $\beta$ reconstruction and a four-dimer completely 
filled trench:  In STM images,    
the filled dangling bond orbitals of the As dimers in the
top layer
extend out far enough that the bright stripes corresponding to the two
As dimers of the mountain appear considerably wider than the dark stripes
corresponding to the two missing As dimers of the trench in the $\beta2$
regions.
It seems quite plausible that the three
dimer structure would appear as if it had a completely filled trench in 
STM pictures, since the filled dangling bond orbitals of the three As dimers
in the top layer could extend far enough out to mask the fourth empty
dimer site.

\begin{figure}[t]
\leavevmode
a) \epsfxsize=7.5cm
\epsffile{clean.eps}

\vspace{0.3cm}

b) \epsfxsize=7.5cm
\epsffile{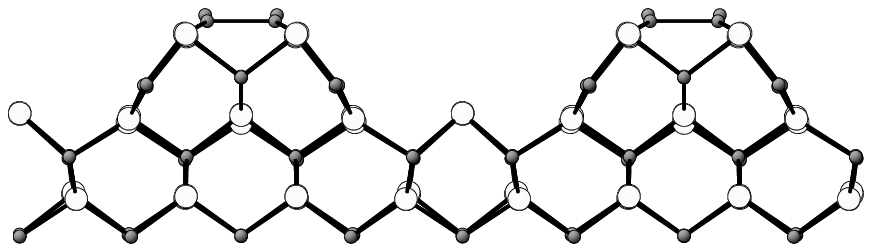}

\vspace{0.3cm}

c) \epsfxsize=7.5cm
\epsffile{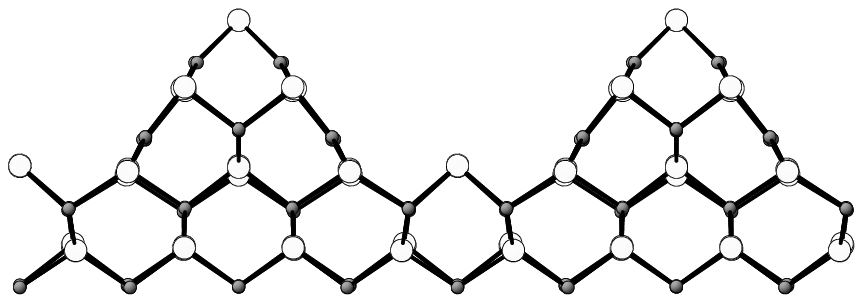}

\vspace{0.3cm}

\caption{\label{layer-scenario} Growth scenario with nucleation of a new layer,
starting from the clean $\beta 2$ surface (a), nucleation of the four-atom Ga cluster
A$_4$A$_2$A$^{''}_4$A$^{''}_2$ (b), and As$_2$ adsorption on the Ga cluster (c).}
\end{figure}

For nucleation of a new layer, we consider the possibility that
the three-Ga atom cluster A$_4$A$_2$A$^{''}_4$ and 
the four-Ga atom cluster A$_4$A$_2$A$^{''}_4$A$^{''}_2$
(see Fig.~\ref{layer-scenario}b) 
could gain in stability by getting "capped" with As$_2$
(Fig.~\ref{layer-scenario}c). 
Similar to adsorption in the trench, we find that an As$_2$ molecule 
with only three backbonds is rather weakly bond, by 1.7 eV, and thus can only
play the role of an intermediary species in growth. 
However, for the four-Ga atom cluster in the top layer, 
we find that As$_2$ binds even more
strongly there than on a cluster of Ga adatoms in the trench.  The binding
energy for an As addimer on this four-atom cluster on the mountain is
2.7 eV.  This indicates that
under the usual growth conditions, any such four-atom cluster which forms
is likely to be rapidly "capped" by an As addimer, becoming a very
stable nucleus for the mountains of the next layer up. Since the Ga adatoms
which fall on any local area are likely to migrate to the trench sites before
such a four-atom cluster on the mountain has a chance to form, until the
nearby Ga sites in the trench are completely filled, we
expect that growth will
generally proceed by a partial filling of the trenches, and formation of local
regions of the $\beta$ structure, followed by nucleation of the mountains of
the next layer up in regions adjacent to locally filled trenches.  This
suggested
growth sequence is shown in Fig.~\ref{growth-scenario}.  
Since nucleation of the
new mountain is expected to proceed before the fourth As dimer adsorbs,
completely filling the trench, we see that the new mountain
must nucleate above the center or above the sloping sides of the mountains
of the original
layer - not above the center of the original trench.  This may explain
why the new mountains are observed to grow above the old mountains
in STM pictures \cite{itoh:98}, instead of
above the trenches.

To make quantitative statements about the importance of arsenic adsorption for
stabilizing structures during growth, we need to take into account the actual
conditions in the growth chamber. In the next section we discuss how this can
be achieved in a growth model.

\pagebreak

\begin{minipage}{7.5cm}
\begin{figure}
\leavevmode
\epsfxsize=7.5cm
\epsffile{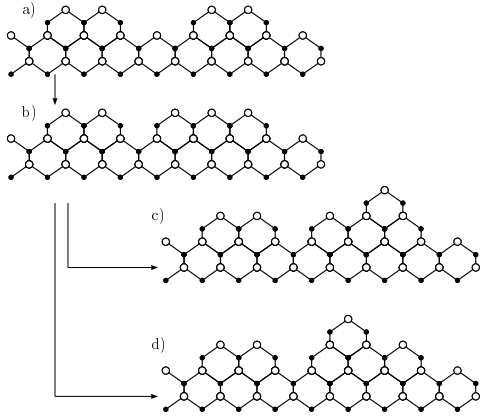}
%
%
\caption{\label{growth-scenario} Growth scenario proposed as a summary of 
the calculated energetics (schematic): the local $\beta$ reconstruction (b) 
acts as a precursor for nucleation of the new layer (c or d).}  
\end{figure}
\end{minipage}

\section{Implications for a growth model}
In empirical growth models for GaAs, often only the Ga atoms are treated explicitely 
\cite{vvedensky:90}. 
This approach is justified if the concentration of As atoms on the growing
surface is completely determined by thermodynamic equilibrium. 
In MBE growth the flux of incoming As molecules is much   
higher (typically by a factor of five) than 
the number of As atoms per unit area and time that are eventually 
incorporated into the crystal (which is equal to Ga flux). Therefore  
the net fluxes of adsorbing and desorbing As$_2$ molecules almost balance, 
and the concentration of As atoms can be considered as being 
close to equilibrium \cite{tersoff:97}. 
In particular, this may be true if the residence time of the arsenic 
molecules is enhanced by a physisorbed precursor state \cite{foxon:77}.
Under the equilibrium assumption, the chemical potential 
of arsenic, $\mu_{\rm As}(p,T)$, is the same both inside the crystal 
and in the gas phase, and is determined by the sample temperature $T$ and 
arsenic partial pressure $p$ in the growth chamber. 
To be specific, we consider growth from As$_2$. The gas phase chemical 
potential can be written down explicitely,
\bea 
\mu_{\rm As}^{\rm As_2}(p,T) &=& -E_{\rm coh}^{\rm As_2} - T S_{\rm vib}(T) \nn
& & - T
S_{\rm rot}(T) - T (S_{\rm trans}(p,T) - 5/2 \; k).
\label{muAs} 
\eea
For the cohesive energy of As$_2$, we employ our calculated result of 2.1
eV/atom, which is in fair agreement with the experimental value of 1.96
eV/atom \cite{hcp}.
The entropies of vibration, rotation and translation are calculated using 
the standard textbook expressions \cite{greiner:87}, with the experimental data on
spectral properties as input data. 

Due to the strong binding of Ga atoms in A$_3$ sites, the evaporation of Ga
adatoms is negligible compared to the incoming flux for temperatures 
below 1000 K.  
In contrast to ref.\cite{tersoff:97}, we therefore argue that the local
concentration of Ga adatoms cannot be described by  
an equilibrium assumption, but should be treated by a kinetic model. 
The incoming flux of Ga atoms drives the system out of equilibrium. 
Only deep in the GaAs bulk, 
a chemical potential for Ga atoms, $\mu_{\rm Ga}^{\rm GaAs \; bulk}$, can be defined by 
the law of mass action for formation of GaAs from the elements,
\be
\mu_{\rm Ga}^{\rm GaAs \; bulk}(p,T) = \mu_{\rm GaAs}(T) - \mu_{\rm As}(p,T).
\label{muGaGaAs}
\ee
Substituting $\mu_{\rm As}(p,T)$ from eq.~\ref{muAs}, this equation 
determines the free 
enthalpy per particle a Ga atom finally reaches after incorporation. 
For various  adsorption
configurations, labelled $i$, the free enthalpy per particle 
$g_{\rm Ga}^{(i)}$ is still well-defined even under non-equilibrium
conditions, 
but no chemical potential can be specified. 
Differences between $g_{\rm Ga}$ for Ga atoms in the beam,
at the surface and in the bulk act as driving force for incorporation  and
growth.  

\begin{figure}[t]
\leavevmode
\epsfysize=6cm
\epsffile{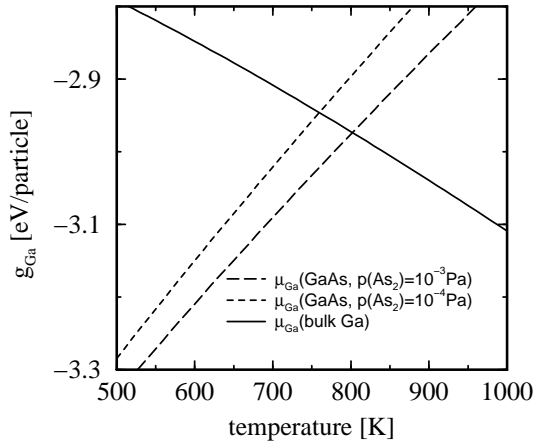}

\vspace{0.3cm}

\caption{\label{mu_vs_T} Free enthalpy per Ga atom in bulk GaAs in
thermodynamic equilibrium with As$_2$ vapor at two pressures, compared to 
the free enthalpy per Ga atom in elemental Ga.}
\end{figure}

In Fig.~\ref{mu_vs_T}, we compare $\mu_{\rm Ga}^{\rm GaAs \; bulk}$ to 
$\mu_{\rm Ga}$ in the bulk of
elemental gallium. By virtue of eq. \ref{muGaGaAs},  
$\mu_{\rm Ga}^{\rm GaAs \; bulk}$ becomes a
function of temperature and As$_2$ background pressure.
For the condensed phases of both GaAs and elemental Ga, we have assumed 
that the pressure dependence of $\mu$ may be neglected. 
The temperature dependence of  $\mu$ for the condensed phase is calculated 
from a Debye model for the lattice 
vibrations,
\bea
\mu_{\rm Ga}(T) &=& -E^{\rm Ga}_{coh} + \int_{0}^{T} \! \! dT' \, c_V^{\rm
Ga}(T') \frac{T'-T}{T'} \\
\mu_{\rm GaAs}(T) &=& -2E^{\rm GaAs}_{coh} + 2 \int_{0}^{T} \!
\! dT' \, c_V^{\rm GaAs}(T') \frac{T'-T}{T'} \\
c_V(T) &=& 9 k_B (T/T_D)^3 \int_{0}^{T_D/T} \! dx \; \frac{x^4 e^x}{(e^x - 1)^2}.  
\eea
Here $E_{\rm coh}$ is the calculated cohesive energy per atom, 
3.1 eV for GaAs and 2.6 eV for Ga,
and $c_{V}(T)$ is the 
specific heat per atom of GaAs and of the bulk phase of Ga, respectively.
The Debye temperature $T_D$ was taken to be 344 K for GaAs \cite{landolt:82}
and 240 K for
Ga \cite{ashcroft:88}.

Fig.~\ref{mu_vs_T} shows that a background pressure of As$_2$ in the 
range of $10^{-4}$ Pa to $10^{-3}$ Pa
is required to stabilize the GaAs crystal under typical growth temperatures, 
between 700 K and 800 K. For $10^{-3}$ Pa As$_2$ pressure, the GaAs surface 
will become unstable against formation of gallium droplets at temperatures 
above \hbox{$~\sim 800$ K.} 

The free enthalpy per atom introduced above allows to compare the 
energetics of structures on the kinetic pathway of growth 
that contain different amounts of gallium and arsenic.
From our {\em ab initio} calculations, we obtain the relative energies
$\Delta e^{(i)}$ of various adsorbate structures on GaAs(001) 
at $p=0$ and $T=0$ with respect to a reservoir of elemental Ga. 
The transfer of a Ga atom from the reservoir into the A$_3$ site on the surface is
endothermic by 0.55 eV. The values of $\Delta e^{(i)}$ can be read from
Tab.~\ref{clusters} using the relation $\Delta e^{(i)}= 0.55 {\rm eV} + \Delta
E^{(i)}/N_{\rm Ga} $, where $N_{\rm Ga}$ is the number of deposited Ga atoms.
Formally, each deposited As atom goes along with the transfer of a Ga atom 
from the Ga bulk to the GaAs bulk reservoir, leading to a free enthalpy change
equal to \mbox{$\mu_{\rm Ga}^{\rm GaAs \; bulk}(p,T) - \mu_{\rm Ga}(T)$}.  
Hence we can directly compute the quantity 
\bea
g_{\rm Ga}^{(i)}& + & \Delta s^{(i)} T = \mu_{\rm Ga}(T) + \Delta e^{(i)} \nn
& + & \bigl( \mu_{\rm Ga}^{\rm GaAs \; bulk}(p,T) - \mu_{\rm Ga}(T)\bigr) N_{\rm As}/N_{\rm Ga},
\label{gGa}
\eea   
with $N_{\rm As}$ the number of deposited As
atoms, and $\Delta s^{(i)}$ being the difference in entropy of a Ga 
atom in adsorption site $i$ and in the Ga bulk. 
This difference contains contributions from vibrational
entropy as well as from configurational entropy, which in turn depends on the
concentration of deposited Ga adatoms on the surface. 

\begin{figure}[t]
\leavevmode
a) \\
\epsfysize=5.0cm
\epsffile{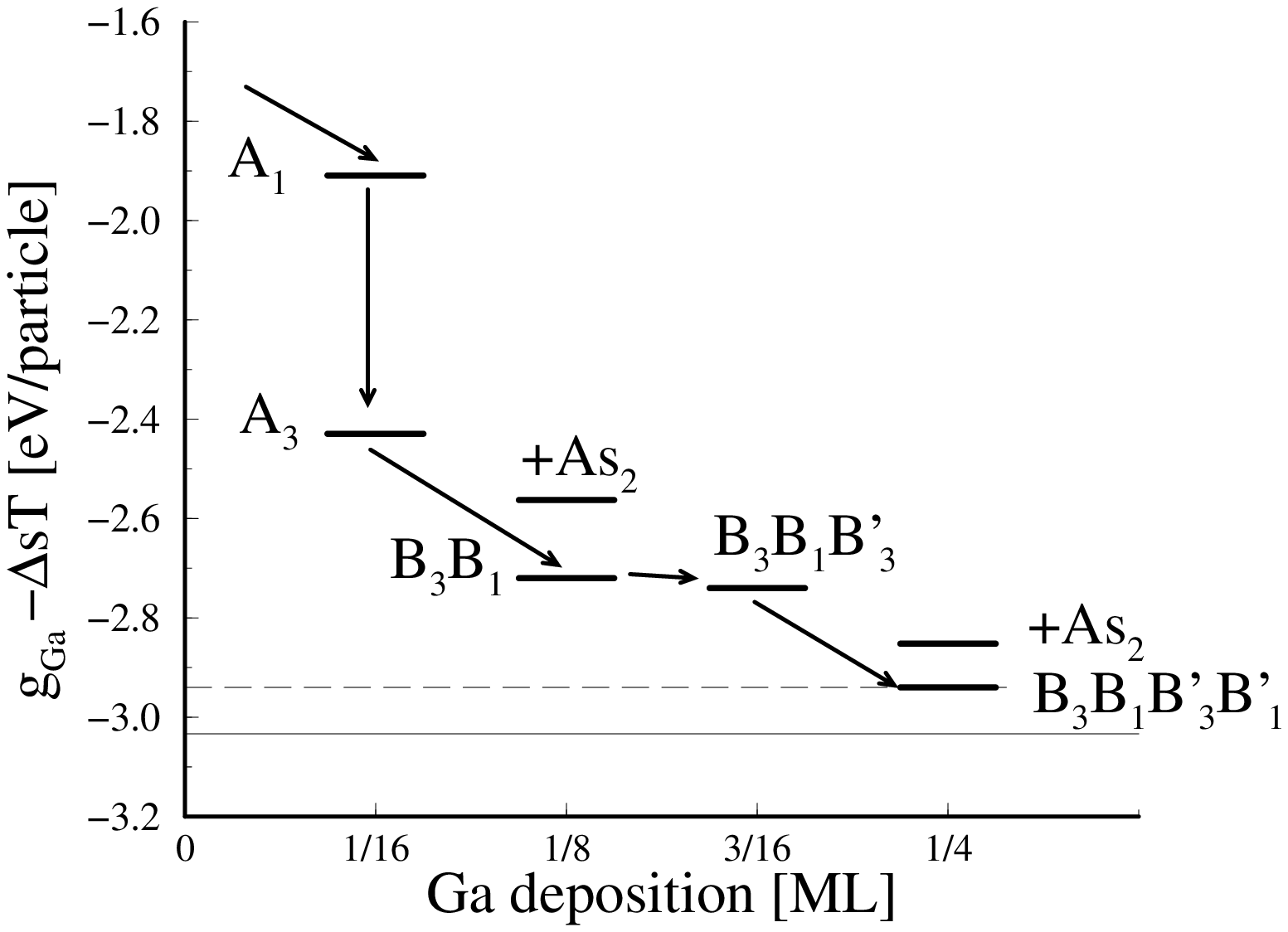}

b) \\
\epsfysize=5.0cm
\epsffile{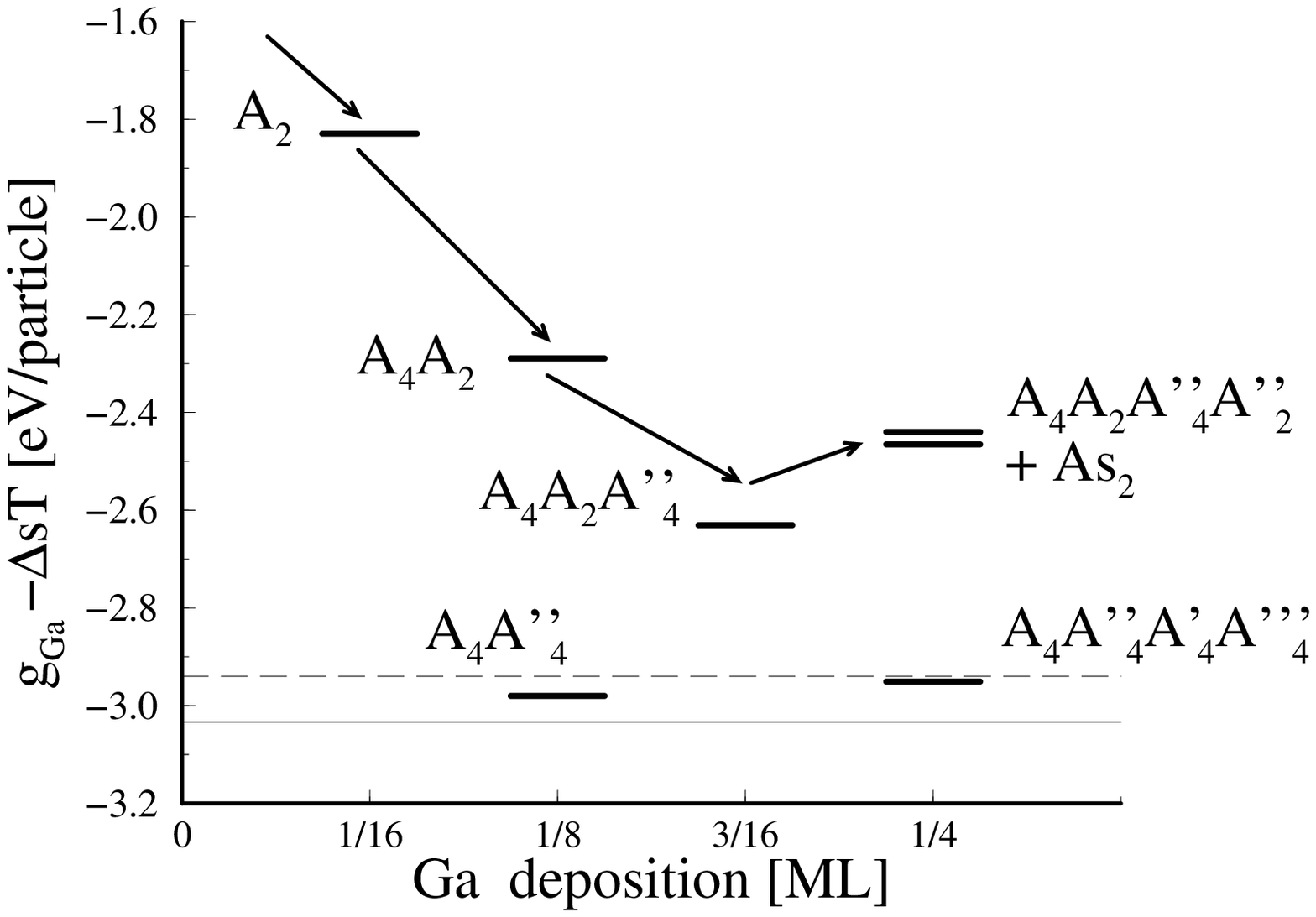}

\vspace{0.3cm}

\caption{\label{casc750} Free enthalpy per Ga atom at $T=750$ K and $p(\rm
As_{2})=10^{-3}$Pa for two kinetic pathways marked by the arrows, a) for filling of the trenches (see also Fig.~\ref{trench-scenario}), and b) for 
nucleation of a new layer (see also Fig.~\ref{layer-scenario}). 
The full and dashed line indicate the chemical potential of a Ga atom in the
GaAs bulk and in elemental Ga, respectively.}
\end{figure}

A gallium atom stemming from the beam source typically runs through several
intermediate configurations on the surface before it becomes part of the GaAs
solid. During this process its free enthalpy $g_{Ga}$ decreases gradually. 
Fig.~\ref{casc750} illustrates the approach of $g_{Ga}$ 
towards its equilibrium value in the GaAs bulk for the two growth 
scenarios discussed above. 
The quantity $g_{\rm Ga}^{(i)} + \Delta s^{(i)} T $ is shown for deposition of
an increasing number of Ga atoms. 
Whenever competing structures with or without As "capping"
are expected, we compare the free enthalpy of both. The values shown in the figure are
calculated for \hbox{$T=750$ K} and \hbox{$p = 10^{-3}$ Pa.} 
Under these conditions,
incorporation of two Ga atoms in the local $\alpha$ structure (B$_3$B$_1$)
has a lower free enthalpy than the local $\beta$ structure (labelled B$_3$B$_1+$As$_2$).
Similarly, larger patches of $\alpha$ structures 
(labelled B$_3$B$_1$B$^{'}_3$B$^{'}_1$) 
have a slightly lower free enthalpy than the corresponding As-capped $\beta$
structures. Thus they will be predominantly uncovered under the considered 
conditions. The situation changes somewhat at temperatures below 700 K or
higher As pressures, 
when As-covered $\beta$ structures become equally probable.
For the structures nucleated on the top layer, 
the calculations show that 
the As$_2$ molecule binds strongly to the four-atom Ga cluster.
Therefore the As-covered structure is preferred 
under a wide range of growth conditions where the 
surface As concentration is in equilibrium with gas phase As$_2$.
However, as can be seen from Fig.~\ref{casc750}, individual Ga dimers
(A$_4$A$^{''}_4$) or Ga dimers above pairs of As dimers, separated by a gap 
(A$_4$A$^{''}_4$A$^{'}_4$A$^{'''}_4$), are still the most stable 
species in absolute terms.

Since the kinetics of Ga incorporation is driven by differences in $g_{\rm Ga}$, a kinetic
model of growth that includes  only the Ga species should describe the stability of
different configurations in terms of the quantities $g_{\rm Ga}^{(i)}$. The
effective parameters of such a model depend on temperature and arsenic
background pressure. Eq.~\ref{gGa} allows to determine them  
from first principles, apart from the yet undetermined entropic contribution $\Delta s T$. 
In a kinetic model,
all microscopic processes are represented by their respective rates. 
In addition to know\-ledge of the (meta-)stable configurations occuring during
growth, a microscopic determination of the rates also requires knowledge about
the transition states. 
Within classical rate theory, the rate for a particular transition is 
given by \mbox{$\Gamma = \exp( ( g^{(i)}-g^{TS} )/kT)$}, with the free enthalpy at the
transition state, $g^{TS}$. It is frequently represented by an activation
energy and a prefactor, $\Gamma = \Gamma_0 \exp(-E_A/kT)$. 
When rates are determined from calculated free enthalpies, the  
contributions to $\Delta s T$ from configurational entropy 
should be omitted, since they are
described implicitely by the number of microscopic processes possible 
in a given situation. 
The vibrational entropies enter the rates only 
through the prefactor, but do not appear in the activation energy. 
Without a detailed calculation of vibrational properties,  
one has to rely on the assumption 
that vibrational contributions to the entropy are similar in 
different configurations. 
Thus they tend to cancel out in $\Delta s$ and 
will not qualitatively change the picture. 
This is equivalent to the assumption of a common  prefactor for all kinetic
processes, as is frequently made in kinetics simulations.

\section{Conclusion}
In conclusion, we have investigated the energetics of two kinetic pathways for
homoepitaxy on GaAs(001) by means of DFT calculations. 
We find that two gallium adatoms on the GaAs(001)
surface interact sufficiently strongly to form stable nuclei. 
One type of nuclei consists of Ga atoms in adjacent three-fold coordinated
sites in the trenches, while another type consists of Ga dimers forming 
on the top layer of arsenic. 
Since single Ga atoms in the top layer are energetically unfavorable and 
constitute an excited state, the formation of Ga dimers in the top layer is
suppressed at low coverages, and growth will be dominated by the nucleation 
of Ga atom pairs in the trenches. 
Filling of the trenches will proceed by further attachment of Ga adatoms to
these nuclei. From our calculations, we expect that islands in the new
layer will preferentially form in those
regions of the surface where locally the $\beta$ reconstruction has formed, 
i. e. where the trenches have been partially filled.
 
Upon further deposition of material, the islands start to grow. 
Quenched STM images have revealed small islands 
which do not yet show the $(2 \times 4)$ pattern \cite{itoh:98}. 
After passing through this metastable
intermediate state, eventually the islands will restructure and display the 
trenches characteristic for the $\beta 2$ reconstruction.

\end{document}